\documentclass[aps,prl,superscriptaddress,twocolumn]{revtex4-1}
\usepackage{graphicx}  
\usepackage{dcolumn}   
\usepackage{amsmath}
\usepackage{braket} 
\usepackage{lineno}
\usepackage{xcolor}
\usepackage{amssymb}

\begin{document}

\title{Strong interband interaction in the excitonic insulator phase of Ta$_2$NiSe$_5$}

\author{Jinwon Lee}
	\affiliation{Center for Artificial Low Dimensional Electronic Systems, Institute for Basic Science (IBS), Pohang 37673, Republic of Korea}
	\affiliation{Department of Physics, Pohang University of Science and Technology, Pohang 37673, Republic of Korea}

\author{Chang-Jong Kang}
	\affiliation{Department of Physics and Astronomy, Rutgers University, Piscataway, New Jersey 08854, USA}
   
\author{Man Jin Eom}
	\affiliation{Department of Physics, Pohang University of Science and Technology, Pohang 37673, Republic of Korea}
  
\author{Jun Sung Kim}
	\affiliation{Center for Artificial Low Dimensional Electronic Systems, Institute for Basic Science (IBS), Pohang 37673, Republic of Korea}
	\affiliation{Department of Physics, Pohang University of Science and Technology, Pohang 37673, Republic of Korea}

\author{Byung Il Min}
	\affiliation{Department of Physics, Pohang University of Science and Technology, Pohang 37673, Republic of Korea}
      
\author{Han Woong Yeom}
	\email{yeom@postech.ac.kr}
	\affiliation{Center for Artificial Low Dimensional Electronic Systems, Institute for Basic Science (IBS), Pohang 37673, Republic of Korea}
	\affiliation{Department of Physics, Pohang University of Science and Technology, Pohang 37673, Republic of Korea}

\date{\today}

\begin{abstract}
    Excitonic insulator (EI) was proposed in 60's as a distinct insulating state originating from pure electronic interaction, but its material realization has been elusive with extremely few material candidates and with only limited evidence such as anomalies in transport properties, band dispersions, or optical transitions.
	We investigate the real-space electronic states of the low temperature phase in Ta$_2$NiSe$_5$ with an atomic resolution to clearly identify the quasiparticle energy gap together with the strong electron-hole band renormalization using scanning tunneling microscopy (STM) and spectroscopy (STS).
	These results are in good agreement with the EI transition scenario in Ta$_2$NiSe$_5$.
	Our spatially-resolved STS data and theoretical calculations reveal further the orbital inversion at band edges, which indicates the exciton condensation close to the Bardeen-Cooper-Schrieffer regime.
\end{abstract}

\maketitle



	Many-body interactions in metallic solids often induce insulating states such as Mott insulators through electron-electron interactions and Peierls insulators through electron-phonon interactions~\cite{Imada}. 
	Excitonic insulator (EI) is another type of interaction-driven insulators formed through a purely electronic mechanism from semimetals or semiconductors with small energy gaps~\cite{Jerome,Halperin}.	
	A valence electron excited to a conduction band leaves a hole in the valence band, and they can pair into an exciton~\cite{Ashcroft}.
	When the carrier concentration and the dielectric constant are unusually small, the hole potential is poorly screened leading to enhanced exciton binding energy greater than the energy gap of the system.
	Then, the spontaneous exciton formation occurs and these bosonic quasiparticles condense into the same ground state.
	This unusual condensate, called the EI, results in flat band edges and an enlarged energy gap~\cite{Jerome}.
	In a semimetal, the electron-hole interaction is relatively weak and the phase transition follows the same way as the condensation of Cooper pairs in Bardeen-Cooper-Schrieffer (BCS) superconductors, while the attractive interaction is strong in a semiconductor and the transition corresponds to the Bose-Einstein condensation of excitons~\cite{Bronold}.



	While the EI idea was conceived as early as 1967 and demonstrated in artificial double layer systems with gate voltages or strong magnetic fields at very low temperature~\cite{Payne,Suprunenko,Du,Zhu}, only very few materials were elusively suggested to fall naturally into the EI ground state so far.
	The first experimental suggestion was TmSe$_{1-x}$Te$_x$~\cite{Neuenschwander, Bucher,Wachter2} and later La-doped SmS~\cite{Wachter}, which showed anomalous increases of electric resistivity under high pressure and low temperature.
	Recently, more detailed discussions were carried out on the temperature-driven transition of $1T$-TiSe$_2$~\cite{DiSalvo2,Yoshida,Pillo,Rossnagel,Kidd,Cercellier,Monney,Monney2,Wezel,Monney3,Cazzaniga,Monney4,Watanabe,Kogar}.
	This case, however, has been heavily debated since the insulating property itself is not observed and the transition apparently involves a charge ordering with a lattice modulation~\cite{DiSalvo2,Yoshida,Pillo,Rossnagel,Cercellier,Monney,Monney2,Monney3,Kidd,Kogar}.
	That is, the EI mechanism has to compete with others such as the Jahn-Teller effect or the charge-density-wave (CDW) formation~\cite{Rossnagel,Kidd}.
	At the center of this debate is the complexity of the band structure, valence-band holes at the center of the Brillouin zone (BZ) and excited electrons in three conduction bands at the BZ boundary~\cite{Monney}, which inevitably involves a non-zero momentum phonon in opening a band gap.                                                                                                                                                                                                                      



\begin{figure}
\includegraphics{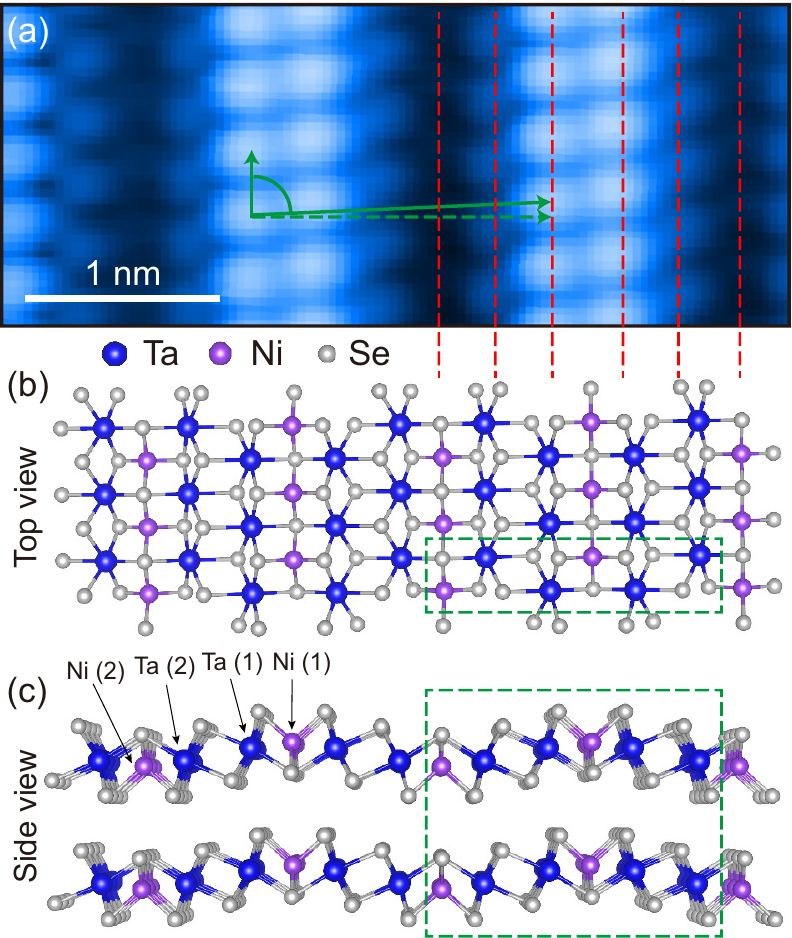}
\caption{\label{Fig01} (Color Online)
STM topography and structure model. (a) Atom-resolved STM topography at 78\,K for Ta$_2$NiSe$_5$ (sample bias $V$\,=\,250\,mV and the set current 300\,pA). The primitive vectors of the monoclinic unit cell are illustrated (green arrows) with deviating slightly from those of the undistorted orthorhombic unit cell (dashed one). (b),(c) Structure model of orthorhombic Ta$_2$NiSe$_5$ with its unit cell in dashed boxes. Red dashed lines connect the atomic positions of the structure model to those of the STM topography.}
\end{figure}

	On the other hand, Ta$_2$NiSe$_5$ was very recently proposed as an EI even at room temperature and ambient pressure~\cite{Wakisaka,Wakisaka2,Kaneko,Kaneko2,Seki,Lu,Larkin}.
	In contrast to $1T$-TiSe$_2$, Ta$_2$NiSe$_5$ has a direct (zero or negative) band gap above the transition temperature $T_c\approx$\,\,326\,K~\cite{Kaneko,Kaneko2} without any CDW involved.  
	The insulating state at low temperature is evident in a transport measurement~\cite{Salvo} and an angle-resolved photoemission spectroscopy (ARPES) work~\cite{Wakisaka,Wakisaka2,Seki}.
	Thus, Ta$_2$NiSe$_5$ has obvious merits to clarify the exciton condensation. Ta$_2$NiSe$_5$ is a layered material~\cite{Sunshine,Salvo} and each layer, as illustrated in Fig. \ref{Fig01}, has two Ta and one Ni chains sandwiched by Se atoms within an orthorhombic unit cell~\cite{Sunshine,Salvo}.
	It undergoes a second-order phase transition to a monoclinic structure at $\sim$\,326\,K with an anomaly in electric resistivity~\cite{Salvo}. 
    ARPES experiments revealed a part of band gap below the Fermi level and its enlargement with the unusual flattening of the valence band edge as the temperature decreases~\cite{Wakisaka,Wakisaka2,Seki}.  
	A model calculation interpreted these observations as the indication of the EI state~\cite{Seki}.
	A recent optical spectroscopy experiment measured the temperature-dependent optical gap of Ta$_2$NiSe$_5$~\cite{Lu,Larkin}.
	Nevertheless, the strong interband interaction, which is the key feature of the EI phase, has not been unveiled yet.
	

	
	In this work, we investigate the real-space electronic states of the EI phase in Ta$_2$NiSe$_5$.
	Atom-resolved local density of states in a real space was obtained using scanning tunneling microscopy (STM) and spectroscopy (STS), which demonstrates the energy gap with sharp peaks at gap edges, corresponding to the flat renormalized band dispersion.
	Moreover, the orbital characters of band edges were inverted in the insulating phase, which is the evidence of the strong electron-hole band interaction.
	The excitonic model calculations revealed that the orbital character inversion also implies the semimetallic band structure in the high temperature phase.
	This leads us to conclude that the EI formation is close to the Bardeen-Cooper-Schrieffer (BCS) regime rather than the Bose-Einstein condensation within the conventional theory of the EI phase.
	

	Single crystals of Ta$_2$NiSe$_5$ were grown using the chemical vapor transport (CVT) method~\cite{Kim}, which were cleaved \textit{in situ} for STM/S measurements.
	STM experiments were conducted using commercial cryogenic STMs (Omicron and Unisoku), for 300 and 78 K, respectively.
	STM topographies were obtained by a constant current mode and a lock-in amplifier was utilized to measure the differential tunneling conductance~$ (dI/dV)$.
	The density functional theory~(DFT) calculations were performed by using the full-potential linearized augmented plane-wave~(FLAPW) band method, as implemented in the WIEN2K package~\cite{Blaha} with the generalized gradient approximation (GGA) for the exchange-correlation~\cite{Perdew}.
	The Brillouin zone integration was done with a 28\,$\times$\,28\,$\times$\,6 k-mesh and the plane-wave cutoff was $R_{MT}K_{max}$\,=\,7.
	The Falicov-Kimball model was used for the exciton model calculation~\cite{Falicov}. 
	

\begin{figure}
\includegraphics{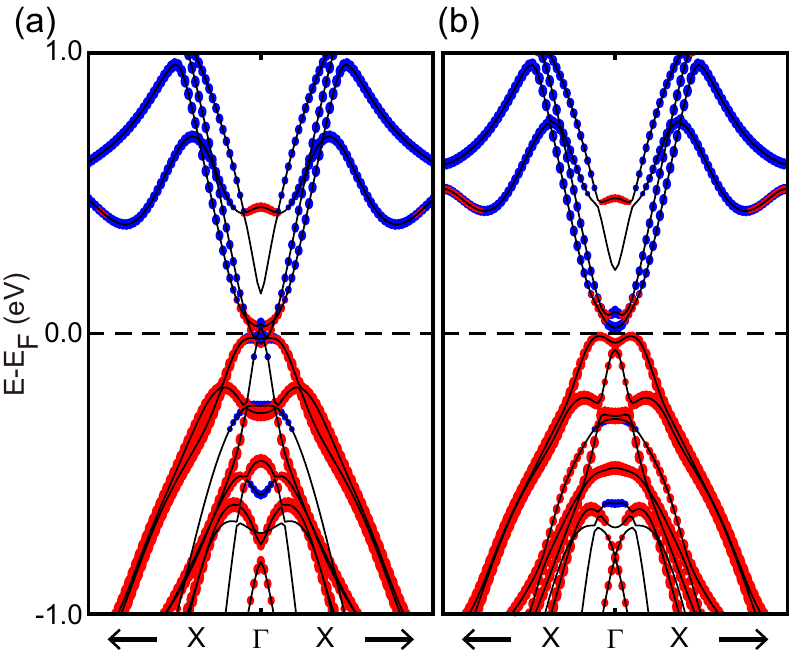}
\caption{\label{str} (Color Online)
Structural phase transition and band structures. (a) Calculated band structure of orthorhombic Ta$_2$NiSe$_5$ (the high temperature phase) with experimental lattice parameters. It is composed of Ta\,5$d$ conduction bands (blue) and Ni\,3$d$ valence bands (red) with a negative band gap. (b) Similar band structure for monoclinic Ta$_2$NiSe$_5$ (the low temperature phase) without any excitonic interaction.}
\end{figure}


	The STM topography taken at 78\,K is shown in Fig.~\ref{Fig01}(a).
	The topography is largely bias-independent within a relevant energy range of $\pm$\,1\,eV, indicating the lack of significant electronic effects such as charge orders or CDW. This is important for the discussion of the transition mechanism.
	The bias-independence implies that the topography is mainly due to the corrugation of the surface Se layer.
	Indeed, the topographic contrast matches well with the corrugation of the structure model (Figs.~\ref{Fig01}(b)~and~\ref{Fig01}(c)) and the x-ray experiment~\cite{Sunshine}. 
	The lattice constants measured by STM are \textit{a}\,=\,3.5\,\AA~and \textit{c}\,=\,15.4\,\AA~with a monoclinic unit cell in good agreement with the x-ray result for the phase below $T_c$.
	Note that this monoclinic structure is a result of the structural transition from an orthorhombic structure at a similar temperature to the electronic transition.
	Nevertheless, the effect of the structural phase transition on the electronic structure is expected to be negligible compared to the band gap formed through the electronic transition~\cite{Wakisaka,Wakisaka2,Seki,Lu}.
	Our own band structure calculations for orthorhombic and monoclinic structures (Fig.~\ref{str}) confirm that the band gap opening due to the structural transition is marginal ($\sim$\,30\,meV)~\cite{supplement}.
	Moreover, the entropy change associated with the transition was found to originate mainly from the electronic structure~\cite{Lu}.
	These results make us focus on the electronic transition.
	

\begin{figure}
\includegraphics{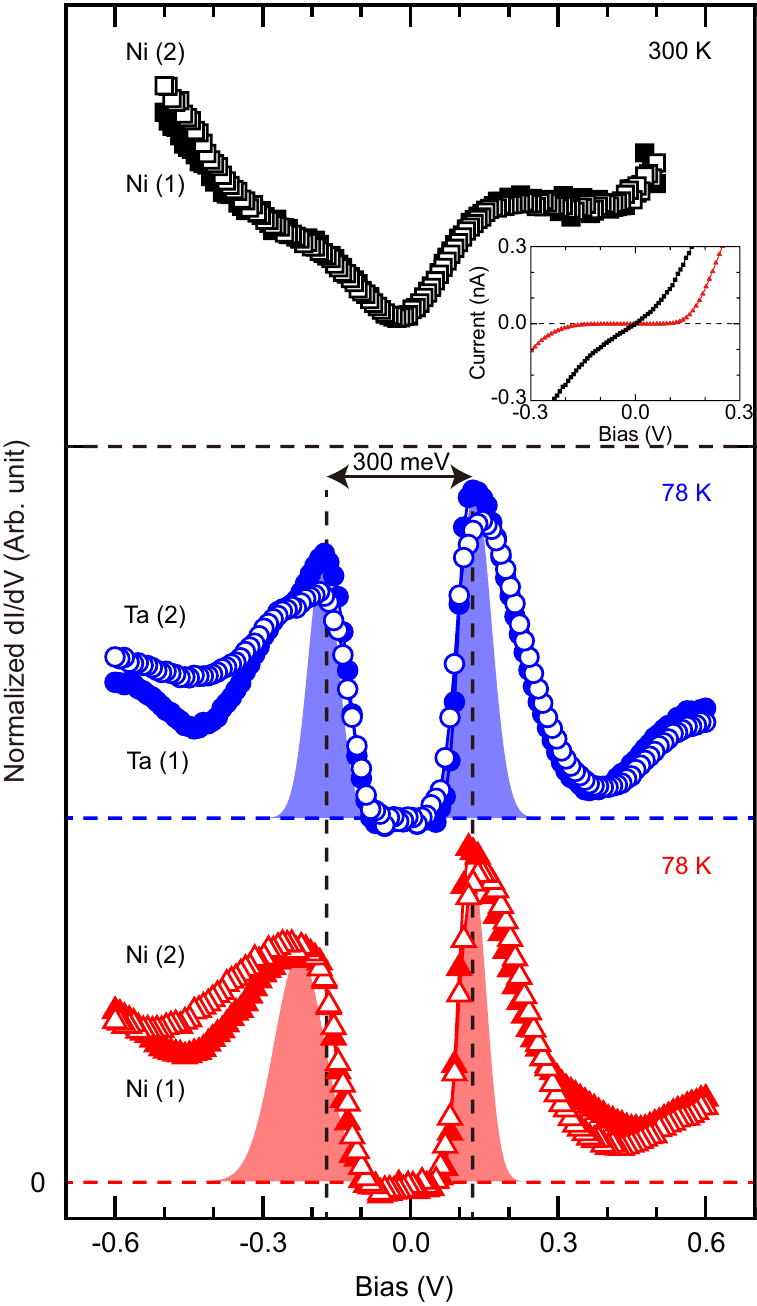}
\caption{\label{Fig02} (Color Online)
Tunneling spectroscopic results at 78\,K and 300\,K. Scanning tunneling spectroscopy $[(dI/dV)/(I/V)]$ data above different Ni (red triangles) and Ta (blue circles) sites (shown in Fig.~\ref{Fig01}) at 78\,K, which are compared with the corresponding data on Ni sites (black squares) at 300\,K. The 300\,K data are uniform over different Ni or Ta sites. The inset shows the tunneling current plot $(I$-$V)$ at 78 (red) and 300\,K (black). The horizontal dashed lines indicate the null spectral intensity. }
\end{figure}


	The electronic phase transition is investigated by STS measurements, $[(dI/dV)/(I/V)]$, which reveals the spatial distribution of local density of states (LDOS) with sub-atomic resolution~\cite{Feenstra,supplement}.
	The room temperature STS data observe finite density of states around the Fermi level although Ta$_2$NiSe$_5$ is reported to have $T_c$ of 325\,K (Fig.~\ref{Fig02}). 
	It obviously reflects the gradual nature of the second-order phase transition with the incomplete gap opening at room temperature together with the thermal broadening of spectral features. 
	In stark contrast, the null density of states near the Fermi level is clear at 78\,K (Fig.~\ref{Fig02}), which evidences an insulating state at low temperature.
	The metal-insulator phase transition is well supported by the tunneling conductance itself shown in the inset.
	The energy gap is as large as 300\,meV with distinct spectral peaks at gap edges.
	
	The DFT calculations (Fig.~\ref{str}) cannot explain this huge energy gap even with the structural distortion, and therefore substantial many-body interactions should be introduced for the electronic phase transition. 
	Before we discuss the interband electron-hole interaction, we consider other types of possible many-body interactions. 
	The on-site Coulomb repulsion (or intraband interaction) can be reasonably excluded; it turns out that the energy gap is still closed in a GGA+U calculation with a well-referenced value of 5\,eV~\cite{Schuster} for the on-site Coulomb repulsion $(U)$ of Ni 3$d$~\cite{supplement}.
	This is because Ta$_2$NiSe$_5$ has large the band width $(w)$, leading to the small value of $U/w$, the order parameter of a Mott transition. 
	On the other hand, the possibility of the electron-phonon interaction has to be considered more carefully. Indeed, the electron-phonon interaction is always present and can compete or cooperate with the electron-hole interaction.
	The previous theoretical study, however, revealed that the electron-phonon interaction cannot open the energy gap solely without the electron-hole interaction but assist the gap opening with a finite electron-hole interaction~\cite{Kaneko2}. Within this theory, the contribution of the electron-phonon interaction on the energy gap can be quantitatively estimated. Using the DFT calculation~\cite{supplement}, we first estimate the phonon momentum $q$ and mode ($\nu$) dependent electron-phonon coupling constants $\lambda_{q\nu}$ at $\Gamma$ point~\footnote{Phonon modes at $\Gamma$ point are associated with the structural phase transition between orthorhombic and monoclinic structures.}. The largest value of $\lambda_{q\nu}$ at $\Gamma$ is 0.1256.
In the strong electron-hole interaction regime~\cite{Kaneko2}, the energy gap enhancement is only a few tens of meV for the largest electron-phonon coupling constant of 0.1256~\footnote{Our value of $\lambda_{q\nu}$ should be divided by the electronic density of states (1.50\,/eV) to be put into this theory.}.
  This gap enhancement is within the energy scale of the relevant phonons~\cite{Larkin2} and the gap size induced by the structural transition discussed above.
Thus, one can conclude that the electronic phase transition is driven mainly by the electron-hole interaction while it can be marginally enhanced by the interaction with phonons. We can thus focus on the EI scenario proposed in this material~\cite{Wakisaka}.

\begin{figure} [t]
\includegraphics{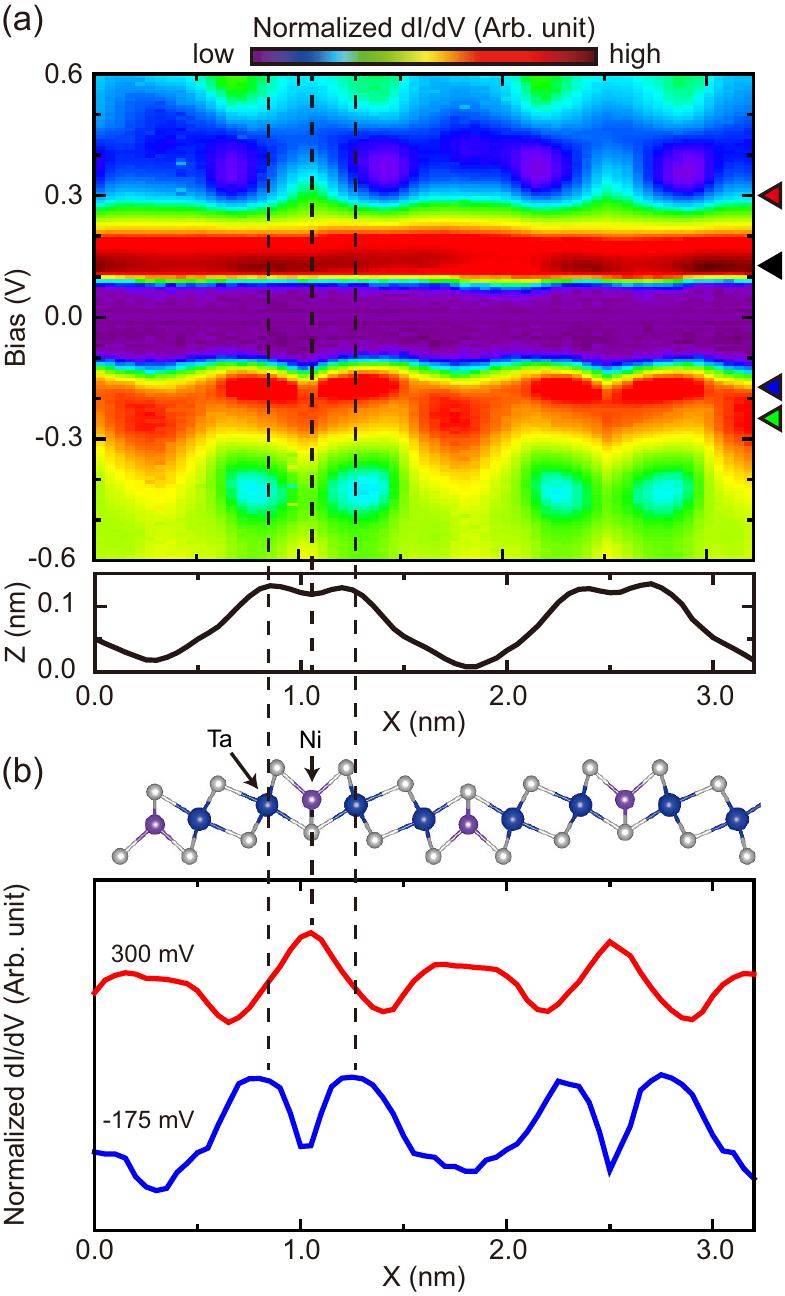}
\caption{\label{Fig03} (Color Online)
Spatial distribution of electronic states at 78\,K. (a) Normalized dI/dV plot of a line STS measurement at 78\,K crossing the Ta and Ni chains (upper) with the topographic profile (lower). (b) Normalized dI/dV profiles with the corresponding structure model at empty (300\,mV) and filled (-\,175\,mV) states. The profile at 300\,mV is vertically shifted for clarification.}
\end{figure}

\begin{figure} [t]
\includegraphics{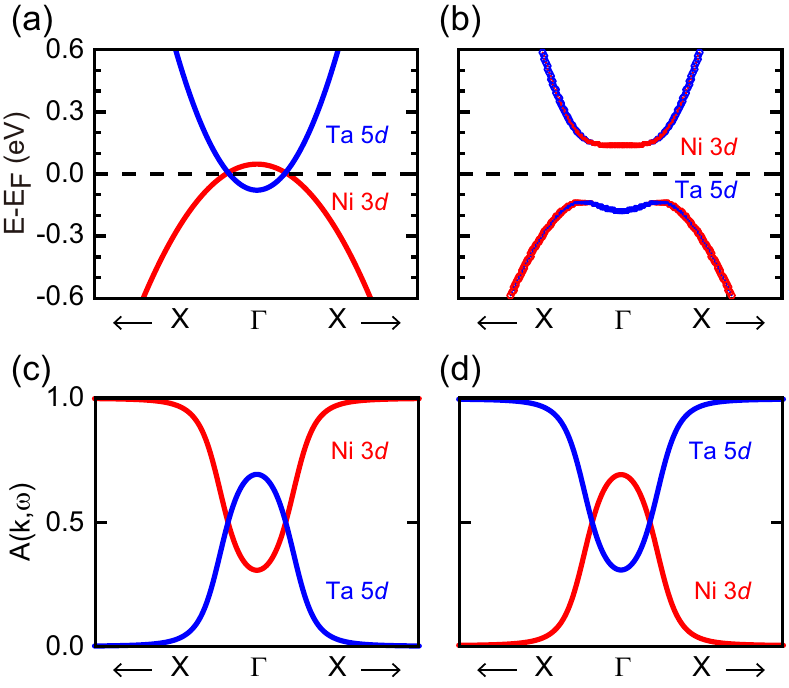}
\caption{\label{Fig04} (Color Online)
Two-band exciton model. Band-gap opening and band character inversion at $\Gamma$, which are induced by the exciton condensation.
(a) Model band structure without the exciton interaction as simplified from the DFT result (Fig.~\ref{str}).
(b) Band structure after including excitonic interaction with an order parameter of $\Delta$\,=\,150\,meV.
(c),(d) Spectral weights of the valence and conduction bands in the excitonic phase of (b).
These weights are also represented by sizes of blue (Ta\,5$d$) and red (Ni\,3$d$) dots in (b).}
\end{figure}

    The spectral features in Fig.~\ref{Fig02} are indeed in good agreement with the EI scenario.
    Strong peaks at the gap edges would correspond to the flat band edges, which is one of the characteristics of the EI phase~\cite{Jerome}.
    This is also consistent with the flat valence band maximum observed in ARPES measurements~\cite{Wakisaka,Wakisaka2,Seki}.
	The asymmetry of the peaks reflects the different band dispersion of those flat band edges in the valence and conduction bands (Fig.~\ref{Fig04}(b)).
	These spectral features are commonly shown in both calculations for the density of states with the excitonic model and our STS spectrum with Feenstra normalization method~\cite{supplement}.
	Detailed discussion on the different spectral shapes is given below with Fig.~\ref{Fig03}.

	Although the optical gap in the low temperature phase of Ta$_2$NiSe$_5$ was measured in the previous research~\cite{Lu}, the single-particle energy spectrum with the energy gap is identified in the present study.
	The reported optical gap ($\sim$\,160\,meV)~\cite{Lu}, which was assigned with nearly zero optical conductivity, seems to correspond to the energy window with zero density of states in our STS measurement.
	Moreover, a broad peak in optical conductivity at 300\,-\,400\,meV seems consistent with the present peak-to-peak energy gap~\cite{Lu}.
	The valence band maximum relative to the Fermi level was measured as -\,175\,meV in the previous ARPES study~\cite{Wakisaka2}, which agrees excellently with the filled-state resonance peak in the STS measurement.
	Thus, the energy gap of the low temperature phase is consistently quantified.
    		

	Beyond the energy gap, the spatial distribution of electronic states is investigated in our STS measurements crossing Ta and Ni chains.
    The differences in point spectra (Fig.~\ref{Fig02}) are also demonstrated in a spatially resolved LDOS map at 78\,K (Fig.~\ref{Fig03}(a)).
	Distinct features are observed around 125($\blacktriangleleft$) and -\,175\,meV($\color{blue}{\blacktriangleleft}$), which correspond to aforementioned edges of valence and conduction bands, and these states are important for the exciton formation.
	Both of them carry satellite features at 300($\color{red}{\blacktriangleleft}$) and -\,250\,meV($\color{green}{\blacktriangleleft}$), respectively, which have distinct spatial distributions as shown in the figure.
	While the feature at 125\,meV($\blacktriangleleft$) has little spatial modulation, that at 300\,meV($\color{red}{\blacktriangleleft}$) is well localized on Ni sites.
	On the other hand, the spectral features at -\,175($\color{blue}{\blacktriangleleft}$) and -250\,meV($\color{green}{\blacktriangleleft}$) are localized at Ta and Ni sites, respectively.
	Note that this localization of each state corresponds to the difference of the spectral intensity in Fig.~\ref{Fig02}.
	This spatial LDOS distribution unveils an important aspect of the gap formation.
	The DFT calculation for Ta$_2$NiSe$_5$ tells us that the bands near the Fermi level are simply composed of parabolic valence and conduction bands from mainly Ni~3$d$ and Ta~5$d$ orbitals, respectively~\cite{Kaneko,Kaneko2}.
	Our own calculation also confirms this simple band structure (Fig.~\ref{str}).
	This contradicts with the STS results where the valence band maximum is localized strongly on Ta and the conduction band minimum has a substantial contribution from Ni atoms.
	Thus, the strong band renormalization has to be involved in the gap-formation transition between Ta and Ni orbitals or conduction and valence band edges.
	

	The energy gap formation is further investigated by model calculations~\cite{Falicov,supplement}.
	As discussed above, the DFT calculation for the non-interacting band structure with the lattice parameters measured by experiments yields a semimetallic phase with a small negative band gap ($\sim$\,-\,50\,meV) as shown in Fig.~\ref{str}(a). 
	In the previous calculation, the conduction (valence) bands were shifted upward (downward) by adding (subtracting) an arbitrary orbital-dependent potential to yield a pre-assumed positive band gap for the high temperature phase~\cite{Kaneko2}.
	However, we cannot find any justification for such an artificial gap opening.
	As the minimal set of a model Hamiltonian, we extract two bands, each from the valence and the conduction band (Fig.~\ref{Fig04}(a)).
	When the coupling of electrons and holes is introduced as a perturbation, the eigenstates near the Fermi level are renormalized with the energy gap opened (Fig.~\ref{Fig04}(b)).
	The gap size depends on the coupling strength ($\Delta$), which is related with the exciton binding energy.
	The renormalized band dispersion at the gap edges is flat together with the substantial hybridization of Ni\,3$d$ and Ta\,5$d$ orbitals.
	The STS measurements are largely reproduced by this two-band model calculation with an order parameter $\Delta=150$\,meV as shown in Fig.~\ref{Fig04}(b).
	That is, the energy gap is opened with a size of 300\,meV ($\approx 2\Delta$) and dominant orbital characters at band edges are inverted due to the exciton formation.
	This inversion is well visualized by the spectral weights of valence (Fig.~\ref{Fig04}(c)) and conduction bands (Fig.~\ref{Fig04}(d)) in the EI phase; Ta\,5$d$ is stronger than Ni\,3$d$ near the $\Gamma$ point in the valence band and vice versa in the conduction band.
	Moreover, we can qualitatively estimate the spatial distribution of the electronic states at the valence band maximum from the model calculation.
	The stronger spectral weight of Ta~5$d$ implies that we have larger density of states on Ta chains than on Ni chains, which is consistent with the normalized $dI/dV$ profile at the valence band edge (-\,175\,mV) in Fig.~\ref{Fig03}(b).

	The major limitation of the simple model is that the experiment observes a relatively well-delocalized state at the conduction band minimum together with the Ni-localized state.
	This might be because Ta$_2$NiSe$_5$ has two degenerate conduction bands and the interaction between these two and the valence band can be more complex than our model.
	Nevertheless, the present result indicates unambiguously that the observed localization of the valence band edge state at Ta chains cannot be explained without the exciton formation.
	Moreover, this observation is in contradiction with the semiconductor-to-EI scenario proposed earlier~\cite{Kaneko2}.
	We performed similar model calculations starting from the semiconducting bare band structure and found no inversion of the dominant band characters at edges~\cite{supplement}.
	In other words, the state at the valence band maximum should have been localized at Ni chains if Ta$_2$NiSe$_5$ were a semiconductor in non-interacting phase.
	That is, the band character inversion is a strong evidence of both the exciton formation and the semimetallic (or~zero~gap) bare band structure at high temperature.
	It is largely consistent with an optical spectroscopy data indicating a zero-gap semiconductor at high temperature~\cite{Lu}.
	
	Another limitation is that our simple model Hamiltonian does not take into account the electron-phonon interaction. As discussed above, while its contribution to the energy gap is marginal, the electron-phonon interaction assists the excitonic condensation~\cite{Kaneko2,Kaneko3}. In addition, a recent study reported the exciton-phonon coupling mode in the low temperature phase~\cite{Werdehausen}, which makes the further investigation on phonons interesting.
	
		
    Summarizing all results, Ta$_2$NiSe$_5$ has a metal-insulator transition from a semimetallic non-interacting phase through the strong interband interaction.
	No other mechanism than EI can plausibly explain these findings together with the experimental data accumulated up to now, such as the band dispersion renormalization.
	Moreover, the present result indicates that the EI phase in this material is close to the BCS mechanism rather than the Bose-Einstein condensation based on the conventional theory~\cite{Halperin,Jerome,Bronold}.
    Within this mechanism, the phases of excitons are not coherent above T$_c$ but, as the temperature is lowered, excitons become gradually coherent and spontaneously condense, opening the band gap.
    It is notable that the condensation starts from a higher temperature than room temperature in contrast to BCS superconductors.
	This is because the binding energy of excitons is two orders of magnitude greater than that of Cooper pairs in BCS superconductors.	
	However, a very recent paper proposed the possibility of a small-gap semiconducting phase above the transition temperature due to preformed excitons in spite of the non-interacting semimetallic bands in Ta$_2$NiSe$_5$~\cite{Sugimoto}.
	Nevertheless, one of our main observations, the orbital character inversion that is the hallmark of a semimetallic non-interacting band, is consistent with this recent paper. The existence of preformed excitons above the transition temperature has to be addressed in future works.
	The further control and manipulation of the exciton condensates are highly promising in Ta$_2$NiSe$_5$ with its extremely high transition temperature.\\
	


This work was supported by the Institute for Basic Science (Grant No. IBS-R014-D1), the KISTI superconducting center (Grant No. KSC-2014-C3-044), and the Global Ph.D. Fellowship Program of National Research Foundation of Korea.


\bibliographystyle{apsrev4-1}
\bibliography{TNS_PRB.bib}

\end{document}


\title{Supplemental Material for\\Strong interband interaction in the excitonic insulator phase of Ta$_2$NiSe$_5$}

\author{Jinwon~Lee}
	\affiliation{Center for Artificial Low Dimensional Electronic Systems, Institute for Basic Science (IBS), Pohang 37673, Republic of Korea}
	\affiliation{Department of Physics, Pohang University of Science and Technology, Pohang 37673, Republic of Korea}
\author{Chang-Jong~Kang}
	\affiliation{Department of Physics and Astronomy, Rutgers University, Piscataway, New Jersey 08854, USA}
\author{Man~Jin~Eom}
	\affiliation{Department of Physics, Pohang University of Science and Technology, Pohang 37673, Republic of Korea}
\author{Jun~Sung~Kim}
	\affiliation{Center for Artificial Low Dimensional Electronic Systems, Institute for Basic Science (IBS), Pohang 37673, Republic of Korea}
	\affiliation{Department of Physics, Pohang University of Science and Technology, Pohang 37673, Republic of Korea}
\author{Byung~Il~Min}
	\affiliation{Department of Physics, Pohang University of Science and Technology, Pohang 37673, Republic of Korea}
\author{Han~Woong~Yeom}
	\affiliation{Center for Artificial Low Dimensional Electronic Systems, Institute for Basic Science (IBS), Pohang 37673, Republic of Korea}
	\affiliation{Department of Physics, Pohang University of Science and Technology, Pohang 37673, Republic of Korea}

\maketitle


\noindent\textbf{Supplementary Note 1: Normalization method of the STS}

Since the tunneling current between the Tip and sample is proportional to $e^{-2 \kappa z}$, where $\kappa$ is related to the work function of the sample and $z$ is the distance between the Tip and sample, the differential conductance $(dI/dV)$ highly depends on Tip-sample distance~\cite{Feenstra}.
Thus, in many cases, $dI/dV$ is not able to represent the density of states properly.
On the other hand, $dI/dV$ divided by the total conductance $(I/V)$, does not have dependence on the Tip and sample distance since we have a factor of $e^{-2 \kappa z}$ both in the denominator and numerator.
Thus, $(dI/dV)/(I/V)$ is proportional to the local density of states~\cite{Stroscio, Feenstra2, Feenstra, Vilan}.

In the normalization procedure, $(dI/dV)/(I/V)$ diverges at $I\rightarrow0$.
To avoid this singularity, normalization with $\sqrt{(I/V)^2+a^2}$, where $a$ is the offset for the total conductance, is used~\cite{Stroscio2, Feenstra}.
At $\left |a \right | \gg \left | I/V \right |$, the normalization factor is approximately constant $\sqrt{(I/V)^2+a^2} \approx \left | a \right |$, which means the normalization not working.
When $\left | a \right |$ is comparable to $\left | I/V \right |$, the peak-to-peak energy gap can depend on the offset $\left | a \right |$.
On the other hand, at $\left | a \right | \ll \left | I/V \right |$, the peak-to-peak energy gap rarely depends on the offset $a$.
Thus, we choose a value of $a$ as small as possible but as a non-zero.
Note that $\left | I/V \right |$ in our measurement is the order of $10^{-10}-10^{-9}\,A/V$ and $2\times10^{-11}\,A/V$ is used for $a$ in our normalization procedure.\\
\\


\noindent\textbf{Supplementary Note 2: Calculations on the electron-phonon coupling constant}

We used the linear response method~\cite{Baroni} implemented in Quantum Espresso~\cite{Giannozzi} for electron-coupling constant calculations. All pseudopotentials used in the calculations were adopted from Standard Solid State Pseudopotentials~\cite{Lejaeghere,Prandini}. To obtain the stable phonon, the fully structural relaxation was performed with the GGA functional until forces exerted on all atoms are smaller than 0.01 $eV/\AA$.\\
\\



\noindent\textbf{Supplementary Note 3: Calculated band structures with lattice distortions}

In our calculation of the band structures for orthorhombic and monoclinic structures in the Fig.~2, the crystal structure for the monoclinic phase is adopted from the experimental structure~\cite{Sunshine}. For the orthorhombic crystal structure, the crystal volume was fixed to the experimental value and the internal atomic parameters were relaxed by using the GGA functional.\\


\noindent\textbf{Supplementary Note 4: Falicov and Kimball exciton model}

In order to examine the excitonic instability properly,
we considered the two-band model with the excitonic interaction term, as follows:
\begin{equation}
\begin{split}
H = &H_{0} + H^{\prime},
\\
H_{0} = &\sum\limits_{\textbf{k}}\varepsilon_{v}(\textbf{k})d_{\textbf{k}}^{\dag}d_{\textbf{k}}
+ \sum\limits_{\textbf{k}}\varepsilon_{c}(\textbf{k})c_{\textbf{k}}^{\dag}c_{\textbf{k}},
\\
H^{\prime} = &\sum\limits_{\textbf{q}}\sum\limits_{\textbf{k}\textbf{k}^{\prime}}
V_{c}(\textbf{q})d_{\textbf{k+q}}^{\dag}d_{\textbf{k}}
c_{\textbf{k}^{\prime}}^{\dag}c_{\textbf{k}^{\prime}+\textbf{q}^{\prime}},
\end{split}
\label{hamiltonian}
\end{equation}
where $V_{c}(\textbf{q}) = 4\pi e^{2}/\epsilon(\textbf{q})q^{2}$ is the direct Coulomb interaction between electrons in the valence and the conduction bands.
For simplicity, we divide the Hamiltonian $H$ into the unperturbed Hamiltonian $H_{0}$, which is the band Hamiltonian extracted from the DFT, and the exciton-induced perturbed Hamiltonian $H^{\prime}$~\cite{Falicov69}.
The first and second terms of $H_{0}$ of Eq.~(\ref{hamiltonian}) are band energies of valence Ni $d$ and conduction Ta $d$ bands, respectively.

Upon cooling, the structural phase transition from orthorhombic to monoclinic structure in Ta$_{2}$NiSe$_{5}$ excludes the charge-density wave (CDW) instability, so that the spanning vector $\vec{w}$ that connects the valence and conduction bands should be zero.
The smallest energy difference between the valence and conduction bands occurs at $\Gamma$, so that we only focus on the Falicov-Kimball exciton model near $\Gamma$.
To apply the above model Hamiltonian to Ta$_{2}$NiSe$_{5}$, we use the perturbation theory with the above unperturbed Hamiltonian $H_{0}$ and the exciton-induced perturbed Hamiltonian $H^{\prime}$.
We approximate the shapes of the valence Ni 3$d$ band and the conduction Ta 5$d$ band which belong to the eigenvalues of the unperturbed Hamiltonian $H_{0}$ as two simple parabolic $d$ bands.

Using the Green function's technique and introducing the excitonic order parameter $\triangle(\textbf{p}) = \Sigma_{\textbf{q}}V_{c}(\textbf{q})\langle c_{\textbf{p}+\textbf{q}}^{\dag}(t)d_{\textbf{p}+\textbf{q}}(t)\rangle$ similar to BCS-like approach, we can obtain one-electron Green's functions for the excitonic insulating phase. For the valence band, one obtains
\begin{equation}
G_{v}(\textbf{p},z)=\Bigg(z-\varepsilon_{v}(\textbf{p})-
\frac{|\Delta(\textbf{p})|^2}{z-\varepsilon_{c}(\textbf{p})}\Bigg)^{-1},
\label{vGreen}
\end{equation}
where $z$ is a (imaginary) frequency and $\varepsilon_{v}(\textbf{p})$ and $\epsilon_{c}(\textbf{p})$ are the valence and conduction band energies for the normal phase, in other words, eigenvalues of $H_{0}$, respectively. For the conduction band, the Green's function $G_{c}(\textbf{p},z)$ is
\begin{equation}
G_{c}(\textbf{p},z)=\Bigg(z-\varepsilon_{c}(\textbf{p})-
\frac{|\Delta(\textbf{p})|^2}{z-\varepsilon_{v}(\textbf{p})}\Bigg)^{-1},
\label{cGreen}
\end{equation}
which has the similar mathematical form with $G_{v}(\textbf{p},z)$.

We can obtain the renormalized valence and conduction bands due to the exciton interaction term in Eq.~(\ref{hamiltonian}) by calculating poles of $G_{v}(\textbf{p},z)$ and $G_{c}(\textbf{p},z)$ in Eqs.~(\ref{vGreen}) and (\ref{cGreen}).
The resulting renormalized valence and conduction bands are
\begin{equation}
\varepsilon_{c,v}^{re}=\frac{\varepsilon_{c}(\textbf{p})+\varepsilon_{v}(\textbf{p})
\pm\sqrt{\big(\varepsilon_{c}(\textbf{p})-\varepsilon_{v}(\textbf{p})\big)^2+4|\Delta(\textbf{p})|^2}}{2},
\label{renor}
\end{equation}
where $\pm$ sign corresponds to the renormalized valence and conduction bands, that is, $\varepsilon_{v}^{re}$ and $\varepsilon_{c}^{re}$, respectively.

The spectral function $A(\textbf{p},\omega)$ is directly proportional to the imaginary part of the Green's function.
For the exciton condensate model, we distinguish the spectral functions of the valence and conduction bands :
\begin{equation}
\begin{split}
A_{v}(\textbf{p},\omega)&=-\frac{1}{\pi}\text{Im}\big[G_{v}(\textbf{p},\omega+i\delta)\big]
\\
&=\frac{\varepsilon_{c}^{re}(\textbf{p})-\varepsilon_{c}(\textbf{p})}
{\varepsilon_{c}^{re}(\textbf{p})-\varepsilon_{v}^{re}(\textbf{p})}
\delta\big(\omega-\varepsilon_{c}^{re}(\textbf{p})\big)
\\
&~~~-\frac{\varepsilon_{v}^{re}(\textbf{p})-\varepsilon_{c}(\textbf{p})}
{\varepsilon_{c}^{re}(\textbf{p})-\varepsilon_{v}^{re}(\textbf{p})}
\delta\big(\omega-\varepsilon_{v}^{re}(\textbf{p})\big),
\label{vSpect}
\end{split}
\end{equation}

\begin{equation}
\begin{split}
A_{c}(\textbf{p},\omega)&=-\frac{1}{\pi}\text{Im}\big[G_{c}(\textbf{p},\omega+i\delta)\big]
\\
&=\frac{\varepsilon_{c}^{re}(\textbf{p})-\varepsilon_{v}(\textbf{p})}
{\varepsilon_{c}^{re}(\textbf{p})-\varepsilon_{v}^{re}(\textbf{p})}
\delta\big(\omega-\varepsilon_{c}^{re}(\textbf{p})\big)
\\
&~~~-\frac{\varepsilon_{v}^{re}(\textbf{p})-\varepsilon_{v}(\textbf{p})}
{\varepsilon_{c}^{re}(\textbf{p})-\varepsilon_{v}^{re}(\textbf{p})}
\delta\big(\omega-\varepsilon_{v}^{re}(\textbf{p})\big).
\label{cSpect}
\end{split}
\end{equation}
Note that integrals over the whole energy range $\omega$ of both $A_{v}(\textbf{p},\omega)$ and $A_{c}(\textbf{p},\omega)$ are exactly 1.

\clearpage

\begin{figure*} [h]
\renewcommand{\figurename}{Supplementary~Figure}
\includegraphics{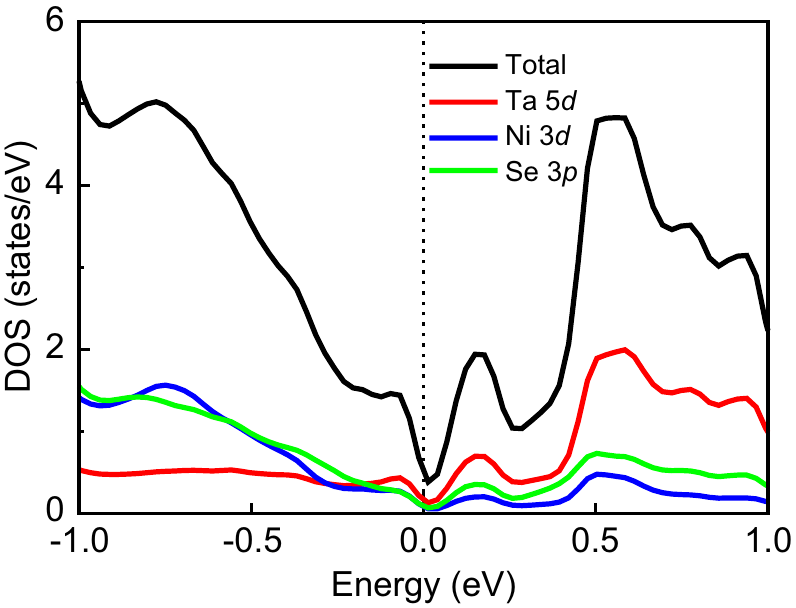}
\caption{ \label{SFig01} (Color online)
Density of states from GGA+U calculation. Note that the on-site Coulomb repulsion does not open the energy gap. The $U$ value of 5\,eV was adopted from that of Ni 3$d$ in NiSe$_2$~\cite{Schuster12} since the valence band top is mainly from Ni.}
\end{figure*}

\clearpage

\begin{figure*} [h]
\renewcommand{\figurename}{Supplementary~Figure}
\includegraphics{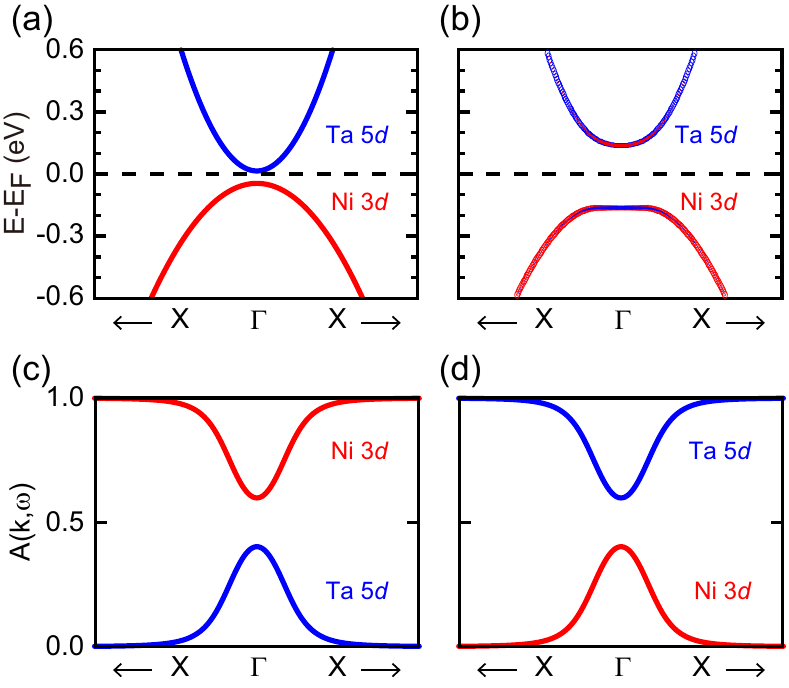}
\caption{ \label{SFig02} (Color online)
Two-band model calculation for the semiconductor to EI phase transition. (a) Semiconducting band structure in a normal phase. (b) Band structure in the excitonic insulator phase with $\Delta$=150 meV. (c) Spectral weights of the valence band in the excitonic phase of (b). (d) Spectral weights of the conduction band in the excitonic phase of (b). Though the excitonic interaction enlarges the energy gap and the effective mass, the band orbital characters are not inverted.}
\end{figure*}

\clearpage

\begin{figure*} [h]
\renewcommand{\figurename}{Supplementary~Figure}
\includegraphics{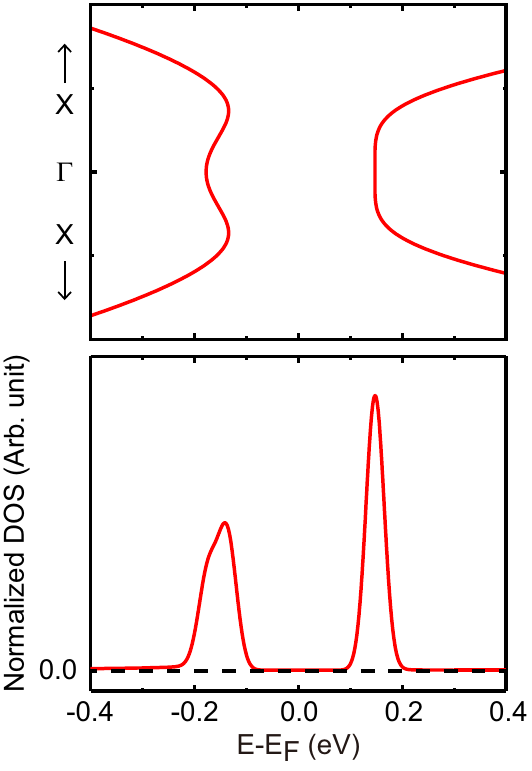}
\caption{ \label{SFig03} (Color online)
Asymmetry in the density of states at band edges obtained from the two-band model. Band structures in the excitonic insulating phase (Fig. 5(b) in the main text) and the density of states are plotted. We apply a Gaussian broadening in the density of states to consider a thermal effect in the experiments. The hat-shape valence band gives a broader and weaker intensity peak in the density of states at the band edge.}
\end{figure*}

\clearpage

\begin{figure*} [h]
\renewcommand{\figurename}{Supplementary~Figure}
\includegraphics{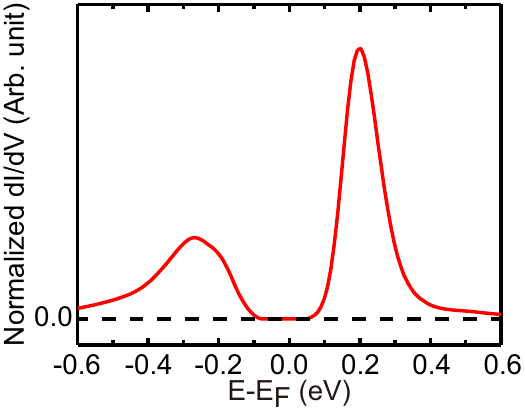}
\caption{ \label{SFig04} (Color online)
Spatially averaged dI/dV with the Feenstra method normalization. The parabolic background in the spectrum is removed by this method. Note that asymmetric peaks in this experimental result are greatly similar to the calculations (Supplementary Figure~\ref{SFig03}).}
\end{figure*}